\documentclass[cameraready]{Interspeech}

\usepackage{multirow}
\usepackage{array}
\usepackage{makecell}
\usepackage{graphicx}
\usepackage{hyperref} 
\usepackage{booktabs}
\usepackage{subcaption}
\usepackage{tabularx}
\usepackage[table]{xcolor}
\usepackage{subcaption}
\usepackage{float}

\title{Exploring the Scale and Diversity of Speech Anti-spoofing Datasets: Experiments and Analysis}

\author[orcid=0009-0008-5771-8976]{Zhuolin}{Yi}
\author[orcid=0009-0001-8465-011X]{Jun}{Xue}
\author[correspondingauthor]{Yanzhen}{Ren}
\author{Yihuan}{Huang}
\author{Yi}{Chai}
\author{Daixian}{Li}
\author{Guanxiang}{Feng}
\author{Jiajun}{Liu}

\address{
	School of Cyber Science and Engineering, Wuhan University, China
}

\email{yizhuolin@whu.edu.cn, junxue@whu.edu.cn, renyz@whu.edu.cn}


\usepackage{comment}

\begin{document}

\maketitle

\begin{abstract}
    The scale of speech anti-spoofing datasets has grown exponentially over the past decade, driven by the assumption that larger data leads to better performance. However, it remains unclear whether indiscriminate scaling commensurately improves model generalization. This study challenges the "scale-first" paradigm by decoupling the impacts of training data scale versus diversity. Through experiments on representative datasets, we report two key findings: (1) Larger is not always better. Expanding data scale excessively under fixed generation methods yields negligible returns and may even degrade cross-domain generalization due to overfitting.(2) Diversity outweighs scale. A smaller composite training set featuring diverse attacks significantly outperforms larger-scale datasets with limited diversity in cross-dataset evaluations. We conclude that future dataset construction should prioritize the diversity of generation methods over scale to effectively enhance model generalization.
\end{abstract}

\section{Introduction}
With the rapid development of speech generation technologies such as TTS and VC, generated speech has achieved remarkable improvements in both naturalness and controllability\cite{xie-etal-2025-towards}. While these advances have facilitated industrial applications and real-world deployments, they have also introduced increasingly severe security threats in areas such as privacy protection and malicious fraud\cite{li_survey_2025}. To address these challenges, the research community has launched a series of ASVspoof challenges\cite{wu_asvspoof_2015, wang_asvspoof_2020, yamagishi_asvspoof_2021, wang_asvspoof_2024}, aiming to effectively distinguish between real and fake speech. However, the accelerating evolution and growing diversity of speech generation methods pose significant challenges to the generalization of fake speech detection, which has compelled researchers to continuously expand the scale of training data to enhance the generalization performance of detection models against unseen attacks.

\begin{figure}[!t]
    \centering
    \begin{subfigure}[b]{0.48\textwidth}
        \centering
        \includegraphics[width=\textwidth, trim=15 55 15 0, clip]{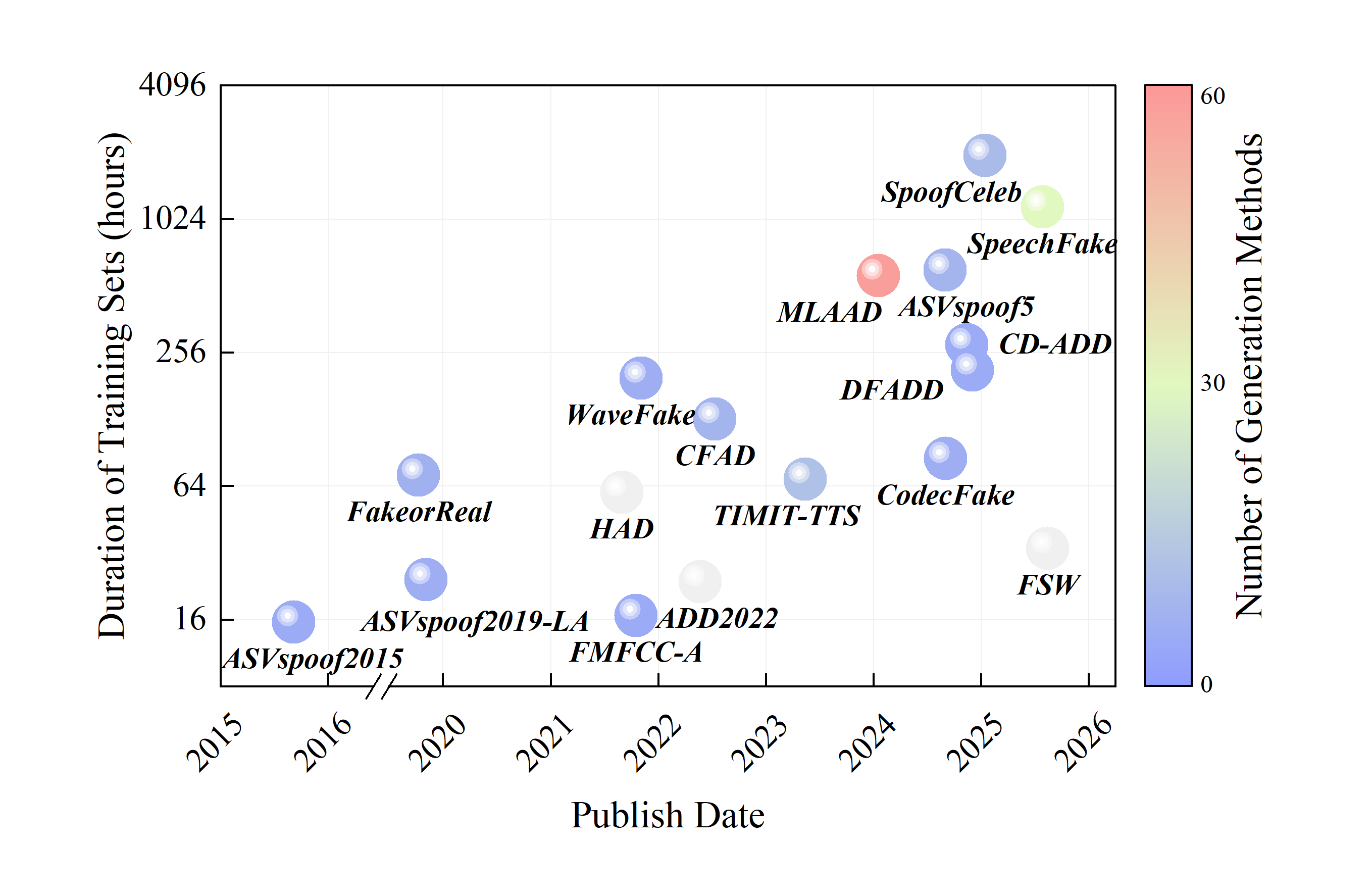}
        \caption{Surveying Speech Anti-spoofing Datasets}
        \vspace{-2pt}
        \label{fig:overview_a}
    \end{subfigure}
    \hspace{1cm}
    \begin{subfigure}[b]{0.45\textwidth}
        \centering
        \includegraphics[width=\textwidth, trim=8 60 12 10, clip]{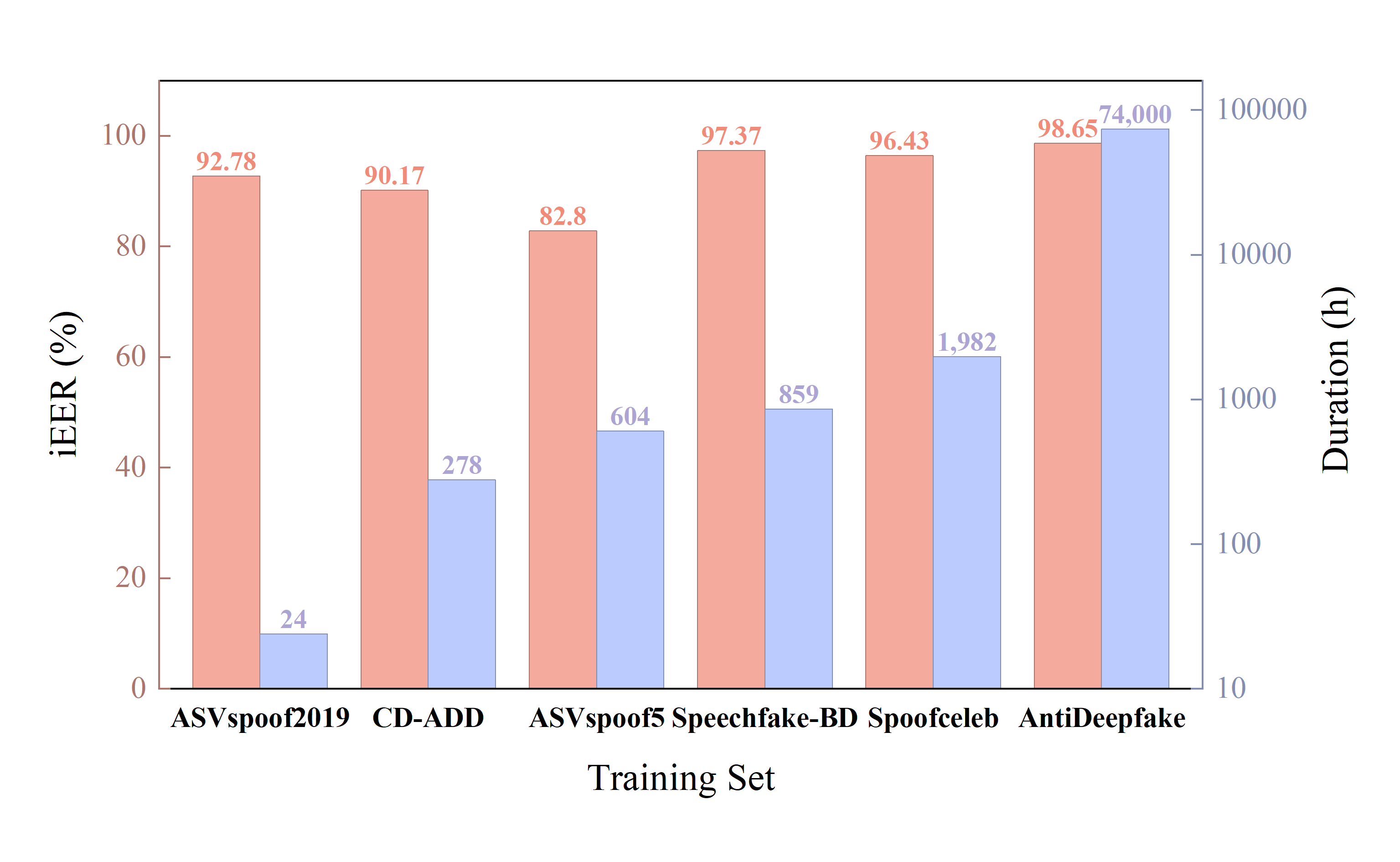}
        \caption{Performance of Models Trained on Datasets with Different Scales}
        \label{fig:intro_fig1b}
    \end{subfigure}
    \caption{Motivations for this work. (a) presents our survey results on the scale and diversity of training sets from representative speech anti-spoofing datasets over the past decade. The vertical axis denotes the total duration of training data, and the color map indicates the number of generation methods covered. For datasets without an explicit data split, statistics are computed over the entire dataset. (b) illustrates the performance evaluation of XLSR\_300M-AASIST models trained on datasets of varying scales, evaluated on the In-the-Wild benchmark (iEER\%, $\uparrow$; iEER = 1 - EER). Results for AntiDeepfake are cited from the source literature, utilizing the XLSR\_1B-FC architecture.}
    \vspace{-10pt}
\end{figure}

We analyze publicly available speech anti-spoofing datasets released over the past decade, with corresponding statistics summarized in Figure~\ref{fig:overview_a}. It can be observed that the scale of training data has exhibited an exponential growth trend over the years. ASVspoof2015\cite{wu_asvspoof_2015}, the first publicly available standardized speech anti-spoofing dataset, contains merely 16 hours of training data covering 5 generation methods. In contrast, the training set of Speechfake\cite{huang_speechfake_2025} encompasses 30 generation methods and spans a total duration of 1,163 hours. AntiDeepfake\cite{ge2025posttrainingdeepfakespeechdetection} further scaled this by utilizing joint training across multiple datasets, aggregating 74,000 hours of training data for the post-training of self-supervised pre-trained models. While achieving superior robustness and generalization, this strategy incurred exorbitant computational costs, requiring eight NVIDIA H100 GPUs for training.

Such a substantial expansion in data scale raises an intuitive question: \textbf{are the costs of constructing and training on large-scale datasets commensurate with the performance gains}? As shown in Figure~\ref{fig:intro_fig1b}, while AntiDeepfake achieves state-of-the-art performance, it yields only a marginal improvement of approximately one percentage point over the runner-up, despite an 86-fold increase in training scale. Notably, Speechfake achieves better generalization performance than Spoofceleb\cite{jung_spoofceleb_2025} despite comprising less than half the training duration. The key distinction appears to lie in the diversity of attacks, as Speechfake covers 30 distinct generation methods, representing three times the number found in Spoofceleb. This leads to the hypothesis that \textbf{indiscriminately increasing data scale under fixed generation methods could be suboptimal, whereas enhancing data diversity represents a more effective strategy to improve generalization}. However, given the inherent discrepancies in speaker distributions, collection pipelines and other confounders across these datasets, such observational findings remain preliminary. Recent work\cite{huang2025data} has investigated the role of data source and generator diversity. In this study, two sets of exploratory experiments are designed to decouple the impacts of confounding factors and conduct a fair comparison.



\begin{figure}[!t]
    \centering
    \begin{subfigure}[b]{0.45\textwidth}
        \centering
        \includegraphics[width=\textwidth]{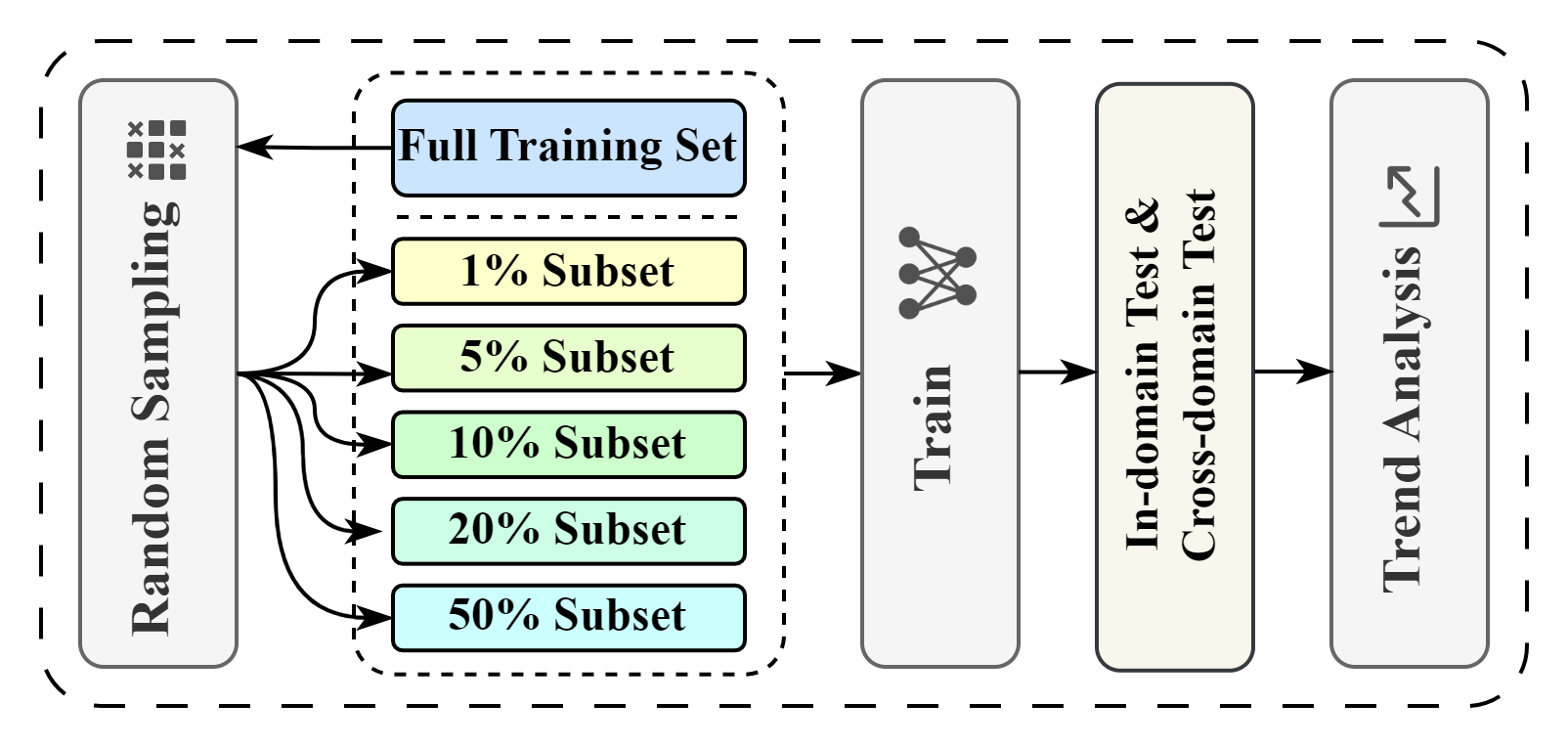}
        \caption{Exploring on Training Data Scale}
        \label{fig:overview_b}
    \end{subfigure}
    \hspace{1cm}
    \begin{subfigure}[b]{0.45\textwidth}
        \centering
        \includegraphics[width=\textwidth]{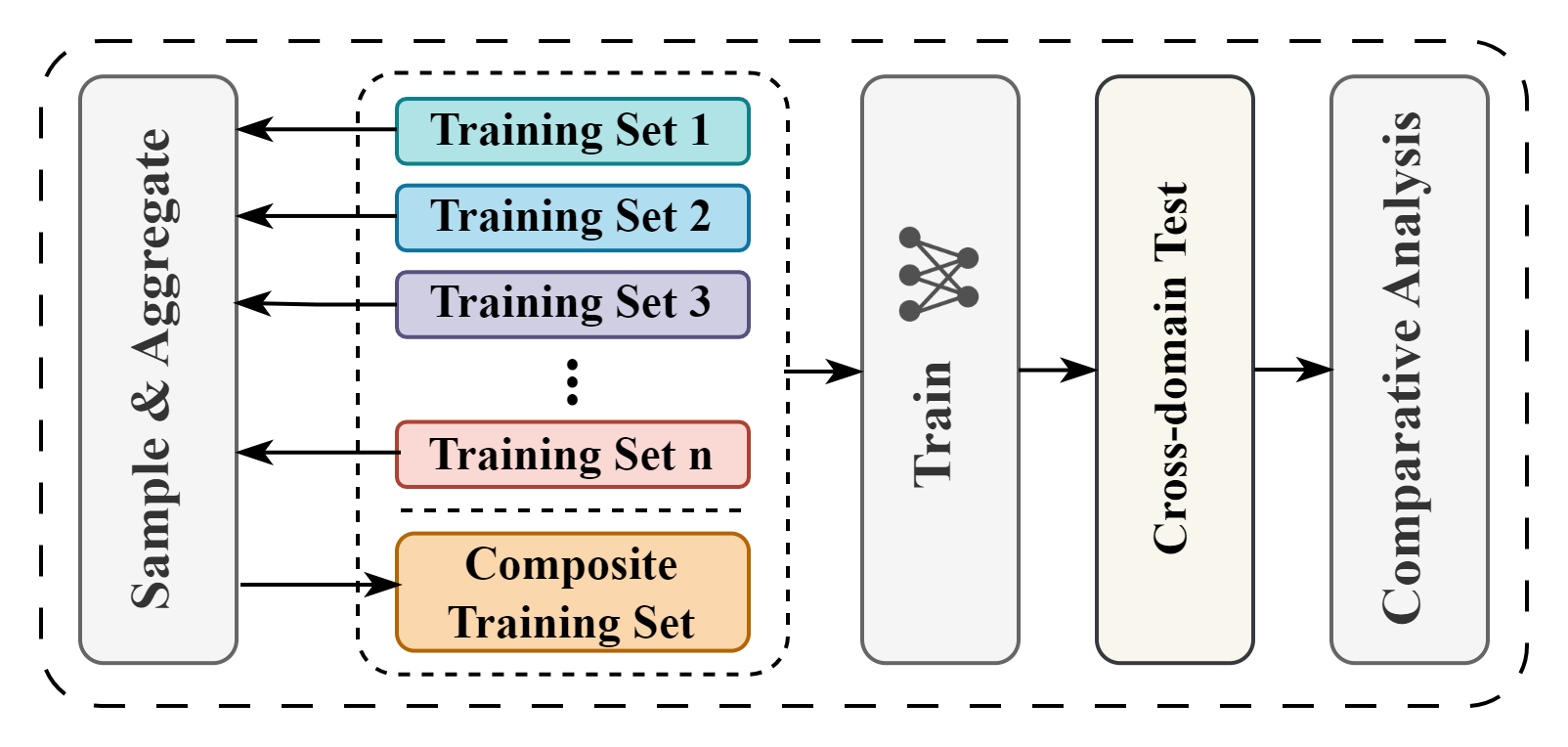}
        \caption{Exploring on Training Data Diversity}
        \label{fig:overview_c}
    \end{subfigure}
    \caption{Overview of our research pipelines. (a) and (b) illustrate the design of our exploratory experiments on training data scale and diversity respectively.}
    \label{fig:overview}
    \vspace{-10pt}
\end{figure}

Figure~\ref{fig:overview} illustrates the detailed pipelines of our exploratory experiments. In the first experiment, training sets are randomly sampled at different proportions to compare the effects of varying data scales on model performance. Subsequently, to investigate the impact of generation methods diversity, we extract a specific number of samples from several distinct datasets to construct a high-diversity composite training set and then evaluate the cross-dataset generalization of models trained on this composite set compared to those trained on the original datasets. Our experimental results reveal the following:
\begin{itemize}
    \item \textbf{Larger is not always better.} As the proportion of training data increases, model performance exhibits an initial improvement followed by a subsequent decline, rather than a continuous positive correlation. This suggests that an excessive sampling from fixed domains may cause the model to overfit to the specific distribution of the training set.
    \item \textbf{Diversity outweighs scale.} Compared to models trained on a single training set, models trained on a more diverse composite training set tend to perform better in cross-dataset evaluations. This indicates that the diversity of generation methods is more vital for improving model generalization ability than the indiscriminate expansion of training data scale.
\end{itemize}

\section{Related Works}

\subsection{Fake Speech Detection}



Fake speech detection aims to distinguish generated speech from real speech. Early research primarily focused on designing discriminative features that capture the differences between the two. Spectral coefficients are utilized by extensive studies\cite{alzantot19_mfcc_cqcc_interspeech,Hamza_MFCC_2022,Borzì_is_synthetic_2022}, while other speech attributes have also been explored to reveal spoofing cues, including F0 subband information\cite{Xue_2022,xue2023learning}, emotion-related features\cite{Conti_emotion_2022}, and breathing-silence correlations\cite{Doan_BTS_2023}. Furthermore, end-to-end methods circumvent the need for handcrafted features by learning representations directly from raw waveforms, with representative works including the RawNet2\cite{Tak_RawNet2_2021}, RawGAT\cite{tak2021RawGAT}, AASIST\cite{jung2022aasist}.

In recent years, self-supervised learning (SSL) models such as Wav2Vec, HuBERT, and WavLM have gained prominence in the speech anti-spoofing field, typically following a paradigm where a pretrained feature extractor is coupled with a downstream classifier. For instance, Wav2Vec-AASIST\cite{tak2022aasist2} replaces the sinc layer of AASIST with a Wav2Vec front-end, and Multi\_SSL\cite{tran_multi_2025} employs multi-kernel gated convolutions as a back-end classifier to capture both local and global speech artifacts. To date, extensive studies have demonstrated that self-supervised pretrained representations significantly enhance generalization performance.

\subsection{Speech Anti-spoofing Datasets}




Over the past decade, the rapid evolution of speech generation methods in both quality and diversity has posed significant generalization challenges, prompting the community to continuously introduce new datasets. Early progress was largely driven by public challenges, most notably the ASVspoof series\cite{wu_asvspoof_2015, wang_asvspoof_2020, yamagishi_asvspoof_2021, wang_asvspoof_2024}, which established standardized benchmarks ranging from early logical and physical access attacks to deepfakes. In addition to the ASVspoof series, FMFCC-A\cite{zhang_fmfcc-_2021} and the ADD\cite{yi_add_2022,yi_add_2023} Challenges filled the gap of Mandarin datasets for fake speech detection.

Complementing these general benchmarks, researchers have proposed various task-specific datasets to address diverse objectives. For detecting speech generated by diffusion models, DiffSSD\cite{bhagtani2025diffssd} and Diffuse or Confuse\cite{Firc_confuse_2024} have been introduced. CodecFake\cite{wu_codecfake_2024} and Codecfake\cite{xie_codecfake_2025} generate fake audio using neural codec models. FamousFigures\cite{Ali_famous_2025} and ZH-FamousFigures\cite{xue2026profiling} target impersonation attacks involving public figures, while FSW\cite{xie_fake_2025} exclusively collects both real and fake speech from real-world social media platforms. Furthermore, datasets such as In-the-Wild\cite{muller_does_2022}, RTCFake\cite{xue2026rtcfake} and VoiceWukong\cite{yan_voicewukong_2024} are designed to evaluate model robustness and generalization in realistic deployment environments.

Recently, speech anti-spoofing datasets have increasingly evolved toward larger scale and higher diversity. Representative datasets include MLAAD\cite{muller_mlaad_2025}, covering 40 languages and 119 TTS models, and Spoofceleb\cite{jung_spoofceleb_2025}, which contains over 2.5 million samples from 1,251 speakers. Speechfake\cite{huang_speechfake_2025} further enlarges these to 46 languages, 40 generation methods and more than 3,000 hours of audio in total. These developments provide a solid data foundation for advancing the generalization of speech anti-spoofing models.

\section{Experimental Setups} \label{sec:exp_setup}
In this section, two complementary sets of experiments are designed to investigate how training data scale and diversity affect the performance of fake speech detection models. While there is no standardized definition of diversity in speech anti-spoofing datasets, it is broadly recognized as a multifaceted concept involving the diversity of generation methods, real data sources, and even speaker variability. For the purposes of this study, we narrow this scope to focus primarily on the variety of generation methods.

\subsection{Data Setups}

\subsubsection{Experiments on Scale}
\textbf{Training Sets.}
We conduct parallel experiments on the training sets of Speechfake-BD and ASVspoof5. Speechfake-BD is a bilingual (Chinese/English) subset of the Speechfake dataset, featuring a large data scale and a relatively wide variety of generation methods. ASVspoof5 is of moderate size and contains exclusively English speech. Both datasets are recent and representative benchmarks in the field, and the statistics of their training sets are summarized in Table~\ref{tab:exp1_train}.

\begin{figure*}[t]
    \centering
    \begin{subfigure}[b]{0.45\textwidth}
        \centering
        \includegraphics[width=\textwidth]{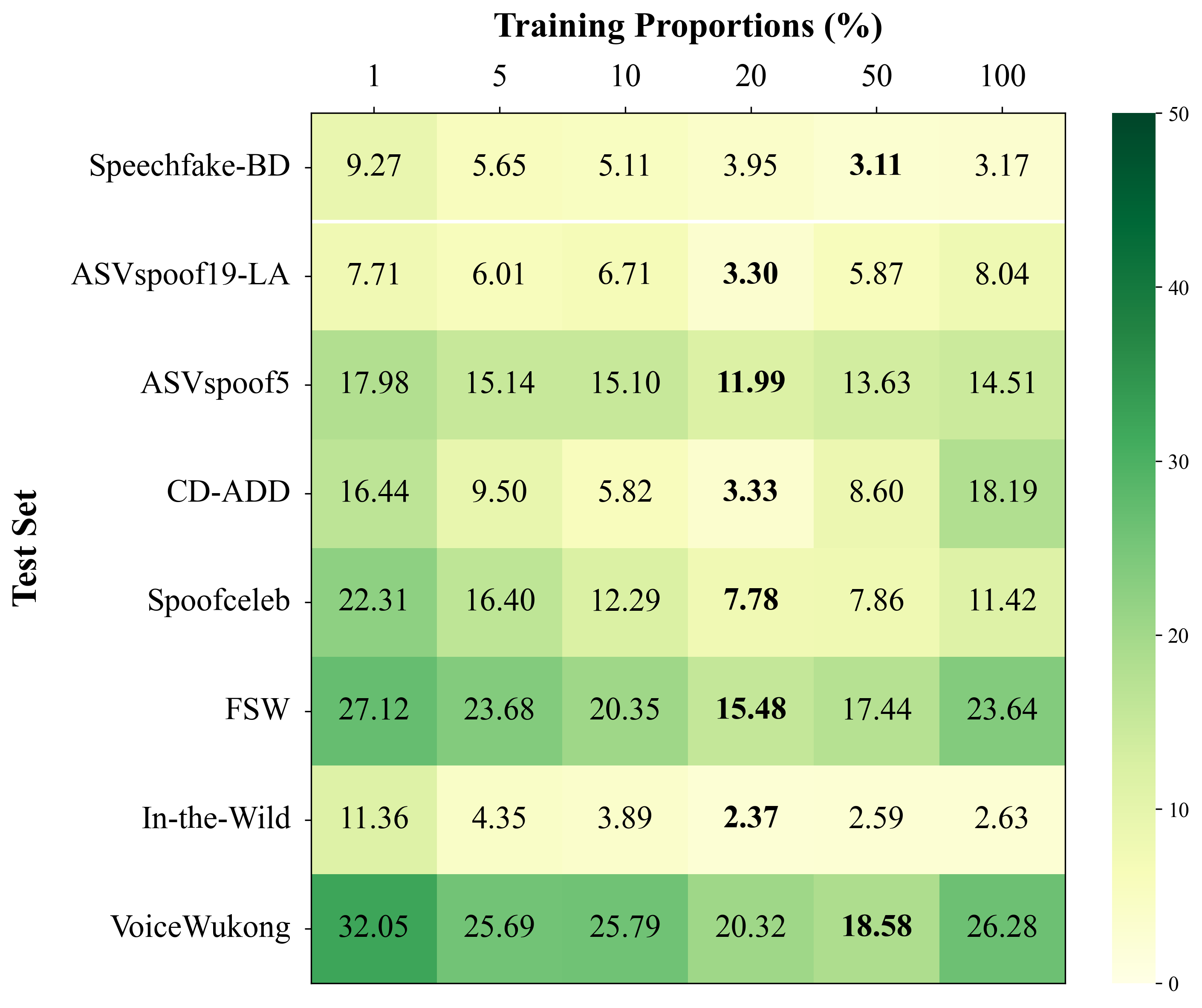}
        \caption{Performance of models trained on varying proportions of Speechfake-BD training set}
        \label{fig:exp_scale_speechfake}
    \end{subfigure}
    \hfill
    \begin{subfigure}[b]{0.45\textwidth}
        \centering
        \includegraphics[width=\textwidth]{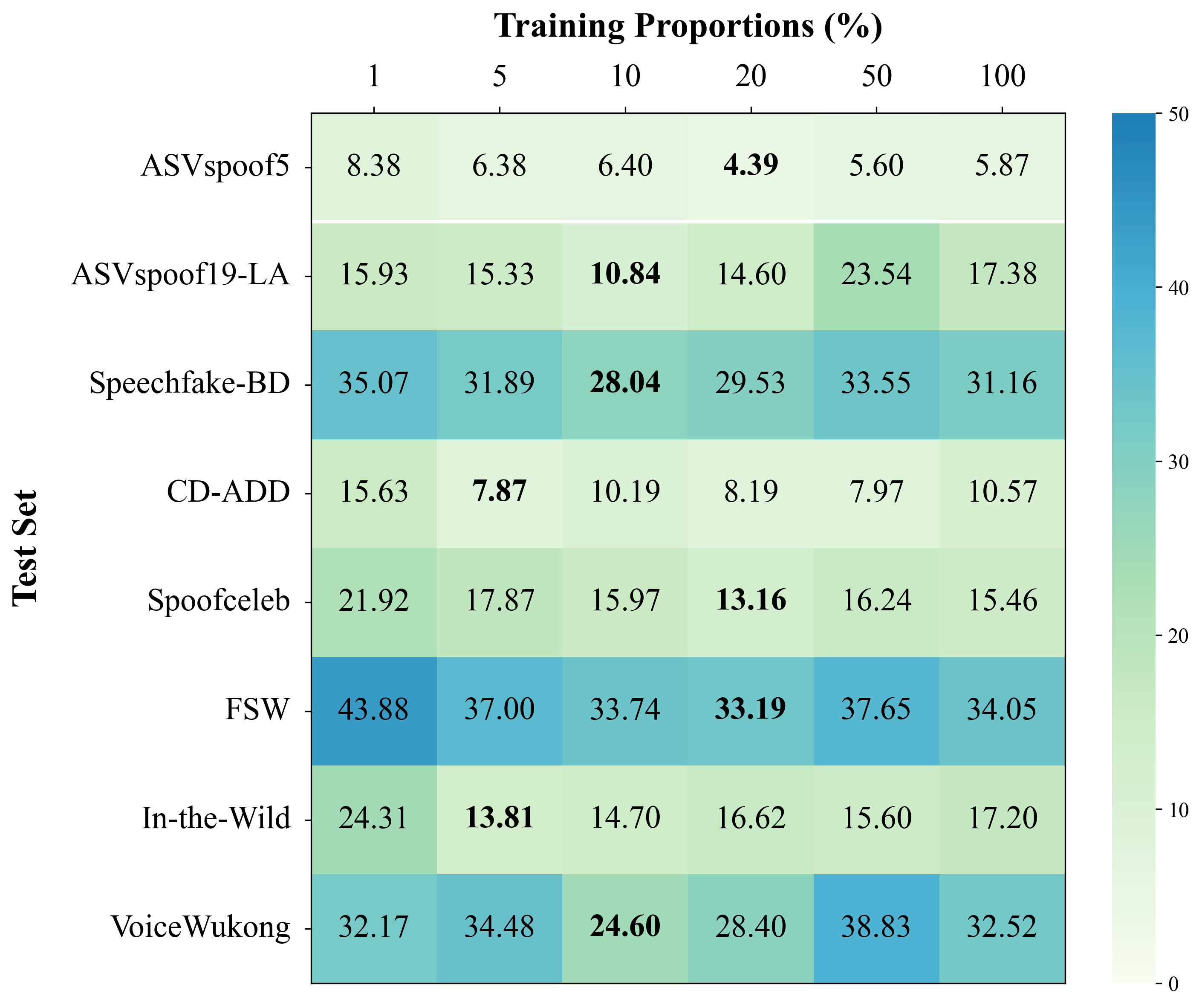}
        \caption{Performance of models trained on varying proportions of ASVspoof5 training set}
        \label{fig:exp_scale_asv5}
    \end{subfigure}
    
    \caption{Evaluation EER (\%, $\downarrow$) of models trained on varying proportions of the (a) Speechfake-BD and (b) ASVspoof5 training sets. The first row in each heatmap shows in-domain test results, while the remaining rows report cross-dataset generalization performance. The best results are highlighted in bold.}
    \label{fig:exp1_results}
    \vspace{-10pt}
\end{figure*}

To investigate the effect of training data scale under a fixed level of diversity, we construct multiple training subsets for each dataset by randomly sampling the training set at proportions of 1\%, 5\%, 10\%, 20\%, 50\%, and 100\% (full set), maintaining the same generation methods as the full set. 

For all experiments, the training of each subset is conducted using the original development set of the corresponding dataset for model selection and training monitoring.

\noindent\textbf{Test Sets.}
For model performance evaluation, we examined both in-domain testing and cross-domain generalization testing. Specifically, for experiments on each dataset, in-domain testing used the test set from the same dataset, while cross-domain generalization testing employed data from other datasets. To more thoroughly validate our findings, in addition to the test sets of Speechfake-BD and ASVspoof5, the cross-domain evaluation was extended to include the test sets of CD-ADD, Spoofceleb, FSW, as well as the full sets of In-the-Wild and VoiceWukong.
 
\subsubsection{Experiments on Diversity}
\textbf{Training Sets.} 
To explore whether diversity plays a more critical role than scale, we construct a new training set by collecting data from four widely used speech anti-spoofing datasets, as shown in Table~\ref{tab:exp1_train}.

For fake speech, we randomly sample 1,000 utterances for each generation method from the training set of each dataset. Although there is limited overlap among generation methods across different datasets, we treat them as distinct methods due to differences in reference speech sources and generation pipelines. All real samples are drawn exclusively from the training set of Speechfake-BD. Detailed statistics of the composite training set are reported in Table~\ref{tab:exp2_train}.

\noindent\textbf{Test Sets.}
Since the composite training set is constructed by sampling and aggregating data from Speechfake, ASVspoof5, Spoofceleb and CD-ADD, it shares overlapping distributions with these datasets. Consequently, evaluations conducted on these four datasets cannot be strictly categorized as either in-domain or cross-domain. To ensure a fair comparison, we therefore perform cross-dataset generalization evaluations only on In-the-Wild, VoiceWukong and the test set of FSW.

\begin{table}[t]
    \centering
    \scriptsize
    \setlength{\tabcolsep}{7.4pt}
    \renewcommand{\arraystretch}{1.2} 
    \caption{Statistics of generation methods and utterances for each training set.}
    \label{tab:exp1_train}
    \begin{tabular}{lcccc}
        \toprule 
        \textbf{Training Set} & \textbf{\#gen} & \textbf{\#fake\_utts} & \textbf{\#real\_utts} & \textbf{\#utts} \\
        \midrule 
        ASVspoof5\cite{wang_asvspoof_2024}  & 8 & 163,560 & 18,797 & 182,357 \\
        Speechfake-BD\cite{huang_speechfake_2025}  & 30 & 629,154 & 75,708 & 704,862 \\ 
        CD-ADD\cite{li_cross-domain_2024}     & 5 & 88,224 & 18,339 & 106,563 \\
        Spoofceleb\cite{jung_spoofceleb_2025} & 10 & 2,309,473 & 230,948 & 2,540,421 \\   
        \bottomrule 
    \end{tabular}

\end{table}
\begin{table}[t]
    \centering
    \caption{Statistics of the composite training set, comprising 63,000 samples and 53 generation methods.}
    \label{tab:exp2_train}
    \scriptsize
    \renewcommand{\arraystretch}{1.2} 

    \setlength{\tabcolsep}{8pt}
    \begin{tabular}{lcccc}
        \toprule
        \textbf{\makecell[l]{Composite\\Training Set}} & \textbf{Source} & \textbf{\#gen} & \textbf{\#utts} & \textbf{\#utts\_total} \\ \midrule
        & ASVspoof5 & 8 & 8,000 &  \\
        & Speechfake-BD & 30 & 30,000 &  \\
        \multirow{-3}{*}{Fake} & CD-ADD & 5 & 5,000 &  \\
        & Spoofceleb & 10 & 10,000 & \multirow{-4.5}{*}{53,000} \\ \midrule
        Real & Speechfake-BD & -- & 10,000 & 10,000 \\ \bottomrule
    \end{tabular}
\end{table}

\subsection{Model Setups}
Although we suspect that both the training data and model capacity may affect model performance, the latter is not the primary focus of this study. Therefore, all experiments are conducted using the fixed model architecture, hyperparameters, and training strategy to eliminate the influence of confounding factors.
Specifically, we adopt the Wav2Vec-AASIST model with the official XLS-R 300M\footnote{https://huggingface.co/facebook/wav2vec2-xls-r-300m} pretrained backbone, and employ RawBoost\cite{Tak_Rawboost_2022} data augmentation during training. This model represents one of the most widely used approaches in current speech anti-spoofing research, making it a reasonable and representative choice for our experimental analysis.

\section{Results and Analysis}
This section presents the experimental results of our exploration on training data scale and diversity across multiple training and test sets, with model performance evaluated using the Equal Error Rate (EER) metric. We conduct a systematic analysis of the results and derive the following two key findings.


\textbf{Finding 1: Model performance does not exhibit a simple positive correlation with the scale of training data.}
Figure~\ref{fig:exp1_results} illustrates the results of the exploratory experiments on training data scale. While varying subsets from an original training set share the same generation methods, increasing the training data scale alone does not yield a monotonic improvement in performance.

For the Speechfake training set, models trained with 50\% and 100\% of the data achieve almost identical performance on the in-domain test set, with the 50\% subset even showing slightly better results. In terms of cross-domain generalization, the best performance is mostly achieved with only 20\% of the training data, after which further increasing the data scale leads to performance degradation. This phenomenon may be attributed to overfitting to the generation methods present in the training data, which weakens the model's ability to detect unseen spoofing attacks. Similarly, for the ASVspoof5 training set, the optimal performance is achieved at the 10\% or 20\% data scale, rather than with the full dataset.

These results suggest that, when generation methods are fixed, simply increasing the amount of training data is not an effective strategy for improving spoofing detection performance. Instead, excessive sampling from the same generation methods provides little benefit for in-domain performance and may even be detrimental to cross-domain generalization and robustness.

\textbf{Finding 2: The diversity of training data outweighs scale in improving model generalization.}
Table~\ref{tab:exp2_result} reports the experimental results obtained using training sets with varying data scales and numbers of generation methods. It can be observed that, across all out-of-domain evaluations, the best performance is consistently achieved by the composite training set, despite its substantially smaller size compared to the other training sets. In particular, the largest training set, Spoofceleb, is significantly larger than Speechfake-BD and the composite training set, yet its generalization performance is still inferior to the latter two, primarily due to the limited diversity in generation methods. This observation leads to the conclusion that, for improving model generalization, diversity contributes more than simply increasing the amount of training data.

\begin{table}[ht]
    \centering
    \scriptsize
    \setlength{\tabcolsep}{3.2pt}
    \renewcommand{\arraystretch}{1.2} 
    \caption{Comparison of cross-domain generalization of models trained on the Composite training set versus full original training sets (EER\%, $\downarrow$). The best results are highlighted in bold.}
    \label{tab:exp2_result}
    \begin{tabular}{lcccccc}
        \toprule
        \multirow{2}{*}[-3pt]{\textbf{Train}} & 
        \multirow{2}{*}[-3pt]{\textbf{dur(h)}} & 
        \multirow{2}{*}[-3pt]{\textbf{\#gen}} & 
        \multicolumn{4}{c}{\textbf{Test}} \\
        \cmidrule(lr){4-7}
        & & & \textbf{Avg} & \textbf{In-the-Wild} & \textbf{VoiceWukong} & \textbf{FSW} \\
        \midrule
        \rowcolor[HTML]{EDF2FB} Composite & 94 & 53 & \textbf{13.03} & \textbf{2.06} & \textbf{19.46} & \textbf{17.58} \\
        CD-ADD     & 278 & 5 & 21.65 & 9.83  & 25.20 & 29.91 \\
        ASVspoof5  & 604 & 8 & 27.92 & 17.20 & 32.52 & 34.05 \\
        Speechfake-BD & 859 & 30 & 17.52 & 2.63  & 26.28 & 23.64 \\
        Spoofceleb & 1982 & 10 & 19.67 & 3.57  & 23.58 & 31.85 \\
        \bottomrule
    \end{tabular}
\end{table}

\section{Discussion}



Although this study analyzes the impact of training data scale and diversity on the generalization performance of speech anti-spoofing models, several aspects remain insufficiently explored and merit further investigation.

Beyond training data factors, model capacity is also an important determinant of generalization performance, as models with higher capacity can learn more complex patterns from the data but may also be prone to overfitting under limited data conditions, thereby directly affecting their ability to generalize to unseen domains. In this work, a unified model with a fixed parameter scale is adopted to eliminate the influence of model capacity variations and focus on the role of training data itself. However, this setting inevitably limits a deeper investigation into the relationship between training data factors and model capacity. Future work could explore different model capacity settings to investigate how model capacity determines the upper limits on the improvements brought by training data scale.

In addition, we primarily rely on random sampling when constructing training subsets with varying scales and levels of diversity, without incorporating more effective sample selection strategies, such as entropy-based sampling methods, which may help reduce data redundancy while improving training efficiency and generalization performance. In particular, given our observation that increasing training data scale does not necessarily lead to performance gains, how to select more informative and complementary training samples under limited data budgets remains an important direction for future research.

\section{Conclusion}

In this paper, we systematically review the evolution of speech anti-spoofing datasets over the past decade and observe that the scale of corresponding training data has exhibited an exponential growth trend. Motivated by this observation, we conduct two exploratory experiments to examine how training data scale and diversity affect model performance. Our results show that scaling data under fixed generation methods does not consistently improve performance and even degrades cross-domain generalization. By contrast, a smaller but more diverse composite training set achieves better generalization than much larger datasets with limited diversity. This study provides practical insights for future dataset construction, highlighting the need to prioritize diversity over merely expanding dataset scale.


\section{Acknowledgements}
This work is supported by the Natural Science Foundation of China (NSFC) under the grant NO.62572358, 62372334

\section{Generative AI Use Disclosure}
Generative AI tools (GPT-4o) were employed exclusively for language refinement and grammatical corrections in this manuscript. The research design, experiments, analysis, and conclusions were conducted and verified entirely by the authors. All authors are responsible and accountable for the work and the content of this paper.

\bibliographystyle{IEEEtran}
\bibliography{mybib}

\end{document}